\documentclass[journal=jacsat,manuscript=article]{achemso}
\usepackage{graphicx}
\usepackage[version=3]{mhchem} 
\usepackage{bm}
\usepackage{color}
\usepackage[latin1]{inputenc}

\newlength\figurewidth
\author{C. Arnold}
\affiliation{Laboratoire de Photonique et Nanostructures, LPN/CNRS, Route de Nozay, 91460 Marcoussis, France}

\author{V. Loo}
\affiliation{Laboratoire de Photonique et Nanostructures, LPN/CNRS, Route de Nozay, 91460 Marcoussis, France}
\altaffiliation{Université Paris Diderot, Département de Physique, 4 rue Elsa Morante, 75013 Paris, France}

\author{A. Lemaître}

\author{I. Sagnes}

\author{O. Krebs}

\author{P. Voisin}

\author{P. Senellart}
\affiliation{Laboratoire de Photonique et Nanostructures, LPN/CNRS, Route de Nozay, 91460 Marcoussis, France}

\author{L. Lanco}
\affiliation{Laboratoire de Photonique et Nanostructures, LPN/CNRS, Route de Nozay, 91460 Marcoussis, France}
\altaffiliation{Université Paris Diderot, Département de Physique, 4 rue Elsa Morante, 75013 Paris, France}
\email{loic.lanco@lpn.cnrs.fr}
\title[\texttt{achemso} Real-time monitoring]
{Cavity-enhanced real-time monitoring of single charge jumps at the microsecond timescale}

\begin{document}

\begin{abstract}

We use fast coherent reflectivity measurements, in a strongly-coupled quantum dot-micropillar device, to monitor in real-time single-charge jumps at the microsecond timescale. Thanks to the strong enhancement of light-matter interaction inside the cavity, the measurement rate is five orders of magnitude faster than with previous experiments of direct single-charge sensing with quantum dots. The monitored transitions, identified at any given time with a less than $0.2\%$ error probability, correspond to a carrier being captured and then released by a single material defect. This high-speed technique opens the way for the real-time monitoring of other rapid single quantum events, such as the quantum jumps of a single spin.

\end{abstract}

\maketitle

Semiconductor quantum dots (QDs) have attracted considerable interest as important building blocks for quantum information experiments based on photon qubits \cite{Shields2007} or stationary spin qubits \cite{Hanson2008}. The environment of the QD, however, is a source of decoherence which can prevent the system from behaving as an ideal two-level qubit. In particular, due to the extreme sensitivity of the QD transition energy with respect to the value of the local electric field \cite{Alen2003}, the random motion of a few charges outside the quantum dot is sufficient to cause a significant broadening of the transition linewidth \cite{Berthelot2006}. This spectral diffusion effect can limit the performances of quantum information protocols relying on indistinguishable single photons \cite{OBrien2009}, coherent control of spin qubits \cite{DeGreve2011}, or resonant excitation of cavity quantum electrodynamics (QED) devices \cite{Fushman2008}. Conversely, this extreme sensitivity provides a tool for fine electric field sensing: Vamivakas \emph{et al.} have recently shown that a QD-based electrometer is sensitive to the motion of a single charge at a distance of a few microns from the quantum dot \cite{Vamivakas2011}.

In the last years, several groups reported the direct observation of slow and discrete spectral fluctuations, related to charges being captured by or escaping from material defects around a quantum dot, at the few seconds or minutes timescale \cite{Empedocles1996,Robinson2000,Besombes2002}. More recently, Houel \emph{et al.} have been able to control charge-by-charge the capture of single carriers, in defects located within $~100$nm of an InAs/GaAs quantum dot \cite{Houel2012}. However, these experiments were all performed with a time resolution of 0.1 second or more, and thus are not well suited to monitor rapid single events occuring at shorter timescales. In a complementary approach, photon correlation experiments have been proposed to extract characteristic spectral diffusion times at the nanosecond timescale \cite{Sallen2010}: this provides information on fast fluctuation processes, but does not allow the direct observation of these fluctuations.

In this Letter we report the real-time monitoring of single charge jumps, with a measurement rate five orders of magnitude faster than for previous experiments of direct single-charge sensing \cite{Empedocles1996,Robinson2000,Besombes2002,Houel2012}. Our technique relies on fast coherent reflection spectroscopy performed on a cavity QED solid-state device. The system under investigation is a deterministically-coupled QD-pillar cavity device in the strong-coupling regime, into which the incident photons are injected with a high input-coupling efficiency \cite{Loo2012}; this ensures that almost all the incident photons will indeed interact with the quantum dot and provide an optical response highly sensitive to the exact value of the QD transition energy. Single events, corresponding to the capture and release of a single charge by a material defect, are distinctly identified with a few microseconds time resolution and with a less than $0.2\%$ error probability. Our measurements reveal a photoinduced acceleration of the charge dynamics, showing that some free carriers are generated in the QD surroundings even under purely resonant excitation. These observations are consistent with a simple model where the capture and release events involve two carriers interacting in the vicinity of the material defect. This technique, beyond the information that it provides regarding the physics of solid-state fluctuations, can be further extended to monitoring other rapid single events such as spin-flips of a single electron or hole, typically occuring between the microsecond and millisecond timescales \cite{Kroutvar2005}.

A single-charge fluctuation in the QD surroundings can significantly shift the QD transition energy, and in the simplest case this transition energy $\omega_{d}$ can take two discrete different values: $\omega_{d}^{(L)}$ if a nearby material defect is loaded by a trapped carrier [Fig. 1(a)], and $\omega_{d}^{(E)}$ if this defect is empty [Fig. 1(b)]. Capture and release of a carrier can then be detected through the modification of the QD spectral response. Here the QD is coupled to an optical microcavity, thus greatly enhancing the device response to these charge fluctuations. The sample consists in a single InGaAs QD inserted in a pillar cavity system; the cavity mode is confined in the growth direction using a GaAs $\lambda$ cavity between two distributed Bragg reflectors, whereas the lateral confinement is obtained thanks to the semiconductor-air refractive index contrast \cite{SupportingInformation}. Spatial and spectral matching between the cavity mode and a selected QD transition have been achieved thanks to the \emph{in-situ} lithography technique \cite{Dousse2008,Dousse2009}. 

The experimental setup is described in Fig. 1(c). The sample is placed inside a helium vapor cryostat, together with a focusing lens and three nanopositioners for the optical alignment.  A continuous-wave monomode laser, with a finely tunable photon energy $\omega$, is focused on and reflected from the micropillar. The incident and reflected powers are measured with fast silicon avalanche photodiodes (APD). Each photodiode is connected to a lock-in amplifier, while the intensity of the incident laser is chopped at 2MHz frequency with an electro-optic modulation system (EOM). This allows the lock-in measurement of both signals with an integration time down to a few microseconds. The setup ensures a near-unity input-coupling efficiency of the incident photons into the micropillar fondamental mode \cite{Arnold2012,Loo2012}, and a very high mechanical stability during tens of hours.

\begin{figure}[ht!]
\centering \includegraphics[width=0.49\textwidth]{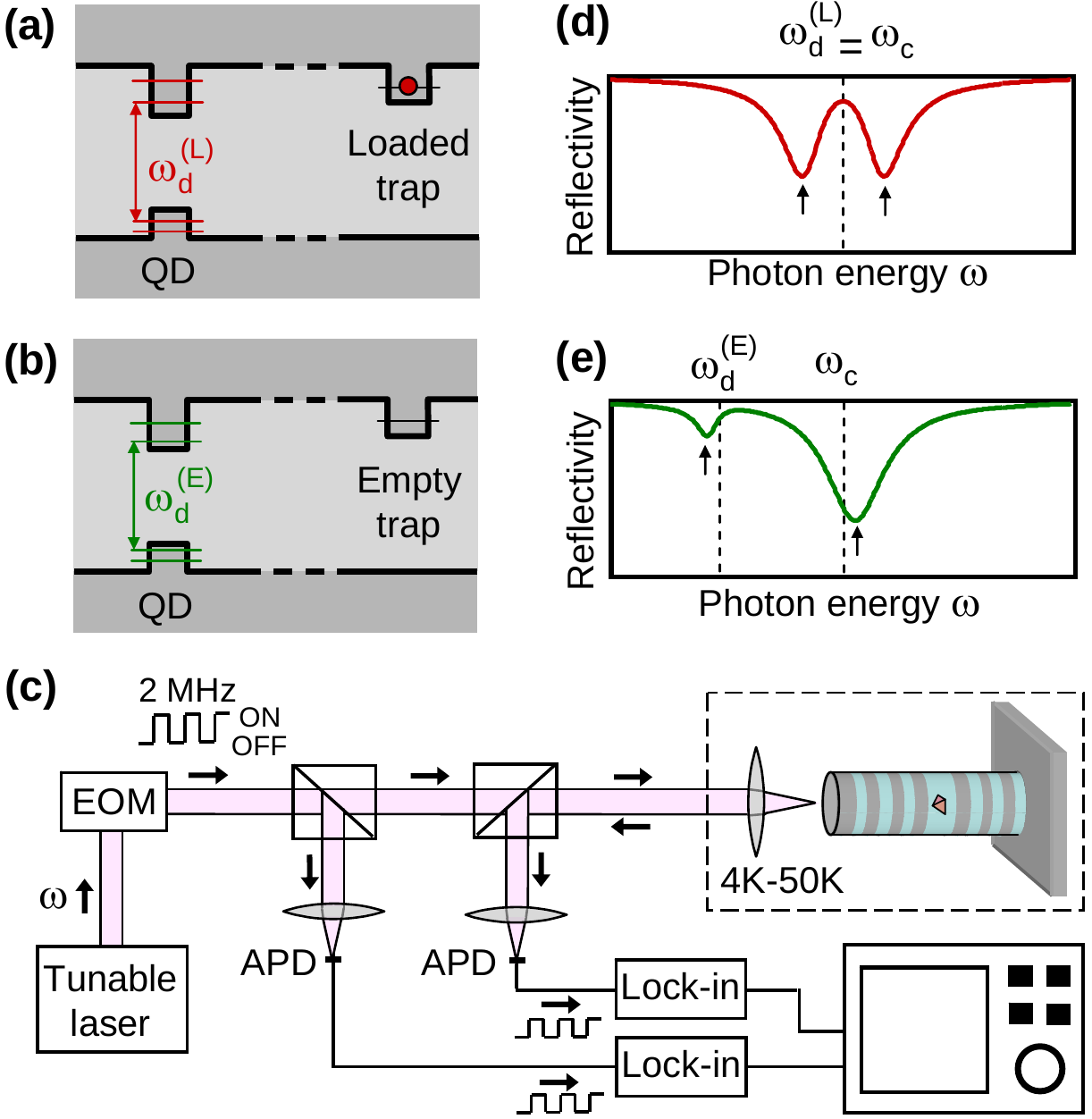}
\caption{(a) and (b) Band structures of an InGaAs QD with a nearby loaded or empty trap, corresponding to QD transition energies $\omega_{d}^{(L)}$ and $\omega_{d}^{(E)}$. (c) Experimental setup. (d) and (e) Typical reflectivity spectra for a loaded and for an empty trap, in the special case where $\omega_{d}^{(L)}=\omega_c$ but $\omega_{d}^{(E)}<\omega_c$. The vertical arrows underline the energies of the mixed exciton-photon eigenstates.  \label{Fig1}}
\end{figure}

In absence of light-matter coupling, a cavity reflectivity spectrum presents only one single Lorentzian dip, at the bare cavity mode resonance energy $\omega_c$. In the strong coupling-regime, on the contrary, the reflectivity spectrum of a QD-cavity device presents two dips located at the eigenenergies of the coupled system \cite{Loo2010}. Fig. 1(d) displays a typical theoretical reflectivity spectrum, computed as a function of the laser photon energy $\omega$, in a configuration where the trap is loaded and where $\omega_{d}^{(L)}=\omega_c$. In such a case both eigenstates have equal photonic and excitonic parts and are symmetrically detuned from $\omega_{d}^{(L)}=\omega_{c}$. When the trap is empty the bare QD transition energy becomes $\omega_{d}^{(E)}$ whereas the bare cavity mode energy $\omega_c$ remains unchanged. As illustrated in Fig. 1(e), if $\omega_{d}^{(E)}$ is lower than $\omega_c$ an assymetrical spectrum is obtained where one eigenstate is mainly exciton-like (energy close to $\omega_{d}^{(E)}$) and the other one mainly photonic-like (energy close to $\omega_c$).

\begin{figure}[ht!]
\centering \includegraphics[width=0.49\textwidth]{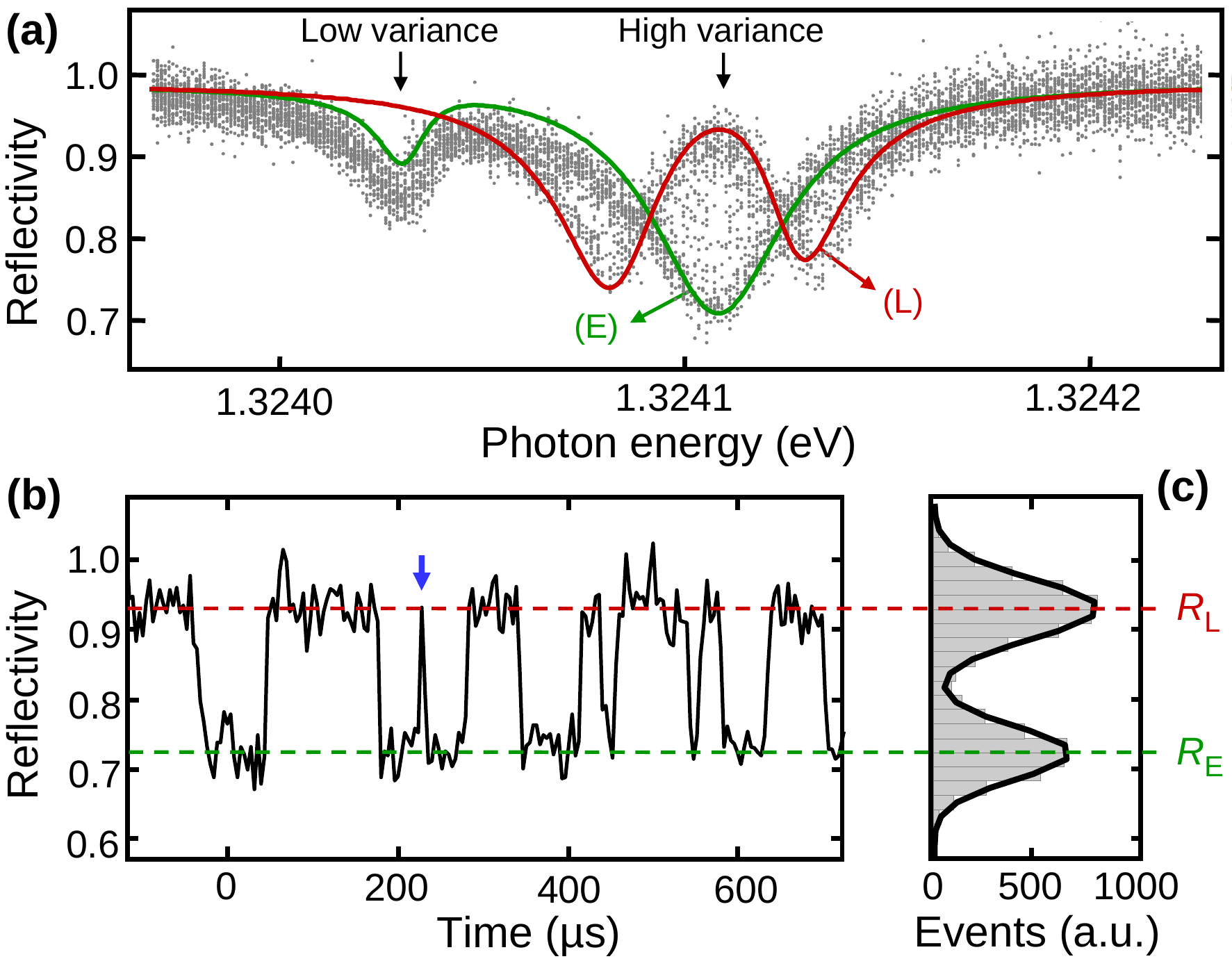}
\caption{\textbf{(a)} Scatter plot of measured reflectivity values vs. photon energy at $T=34.1\mathrm{K}$ and $P_0=1.5~\mathrm{nW}$. Solid lines: theoretical fit. \textbf{(b)} Real-time reflectivity signal at $T=34.1~\mathrm{K}$, $P_0=1.5~\mathrm{nW}$, and $\omega=1.32411~\mathrm{eV}$. Dashed horizontal lines are guides to the eye indicating the two states with reflectivities $R_L$ and $R_E$. \textbf{(c)} Histogram of the reflectivity values measured in the experimental conditions of Fig. 2(b). Solid line: numerical fit with a sum of two gaussian distribution functions centered around $R_L$ and $R_E$. \label{Fig2}}
\end{figure}

Fig. 2(a) presents a scatter plot of several reflectivity values measured as a function of the photon energy $\omega$ for an incident power $P_0=1.5~\mathrm{nW}$. The temperature is $T=34.1~\mathrm{K}$, at spectral resonance between the cavity mode and the QD when the trap is loaded. This plot consists in numerous reflectivity measurements performed with a $50~\mathrm{µs}$ integration time: it highlights the presence of strong reflectivity fluctuations which cannot be accounted for by the experimental noise. The solid curves in this figure are numerical fits obtained with a single set of device parameters \cite{SupportingInformation}, only differing by the values of $\omega_{d}^{(E)}$ and $\omega_{d}^{(L)}$ as in Fig.  1(d) and  1(e). The fitted energy difference, $\omega_{d}^{(E)}-\omega_{d}^{(L)}=75\pm5~\mathrm{µeV}$, is too low to be compatible with a charge fluctuation in the quantum dot itself, but is compatible with a nearby material defect, at a distance of several tens of nanometers from the quantum dot, randomly capturing and releasing single charges \cite{Vamivakas2011,Houel2012}.  The fact that different reflectivity variances are observed in different regions of the spectrum will be discussed below. 

To perform the real-time monitoring of capture and release events, the best experimental configuration is when the photon energy equals $\omega=1.32411~\mathrm{µeV}$, corresponding to the region of very high variance highlighted in Fig. 2(a). Fig. 2(b) displays a typical real-time reflectivity measurement in this configuration. Even with an integration time as short as a few $\mathrm{µs}$, the signal-to-noise ratio is high enough to allow the direct observation of random jumps between two reflectivity values, $R_E$ and $R_L$. These jumps are observed each time the system experiences an $E \rightarrow L$ transition (capture) or a $L \rightarrow E$ transition (release). As an example, the narrow reflectivity peak emphasized in Fig. 2(b) corresponds to the capture and, approximately  $6 \mathrm{µs}$ later, to the consecutive release of a single charge. 

The clear distinction between the loaded and empty states is also illustrated in the reflectivity histogram of Fig. 2(c), well-reproduced numerically with a sum of two gaussian distribution functions centered at $R_E=0.72$ and $R_L=0.93$. The overlap between the two distributions is small enough to allow us, at any moment and with a less than $0.2\%$ error probability, to identify if the system is in the state $E$ or $L$. Furthermore, we can measure with a few $\mathrm{µs}$ precision the time  at which the system undergoes an $L \rightarrow E$ or an $E \rightarrow L$ transition, and thus the time spent in the state $E$, or $L$, between two consecutive transitions.

Fig. 3(a) displays, for different excitation powers, experimental histograms of the time spent by the system in state $E$ before experiencing an $E \rightarrow L$ transition. In each case the probability that the systems remains in state $E$ exponentially decreases with the elapsed time; the corresponding monoexponential law is characterized by a negative slope, whose absolute value gives the transition rate from the empty state to the loaded one, denoted $\Gamma_{E \rightarrow L}$. The four histograms in Fig. 3(a) have been acquired for various values of the excitation power $P_0$, showing that $\Gamma_{E \rightarrow L}$ increases with the incident power. The same set of real-time reflectivity measurements has been used to extract similar histograms (not shown) of the time spent in the loaded state $L$, before experiencing a $L\rightarrow E$ transition: monoexponential decreases are also observed, with a transition rate denoted $\Gamma_{L \rightarrow E}$ that also increases with the incident power.

The relevant quantity, for the analysis of this photoinduced acceleration of the electrostatic fluctuations, is not the incident power but the number $n$ of intracavity photons, which have a non-zero probability of being absorbed, and exciting an electron-hole pair in the InAs wetting layer or in the GaAs barriers. The generated carriers can then participate to the $E \rightarrow L$ transition, if the system is initially in state $E$, or to the $L \rightarrow E$ transition if the system is initially in state $L$. The intracavity photon number $n$ takes the form $n \propto P_0 |1-r_m|^2$, where $r_m$ is the mode reflection coefficient \cite{SupportingInformation}; $r_m$ and thus $n$ take different values if the device is in state $E$ or $L$. Fig. 3(b) shows the transition rates $\Gamma_{E \rightarrow L}$ and $\Gamma_{L \rightarrow E}$, as a function of $n$. For a real-time measurement at a given incident power $P_0$, the value of $n$ considered is the one in state $E$ for the $E \rightarrow L$ transition, and the one in state $L$ for the $L \rightarrow E$  transition.

\begin{figure}[ht!]
\centering \includegraphics[width=0.49\textwidth]{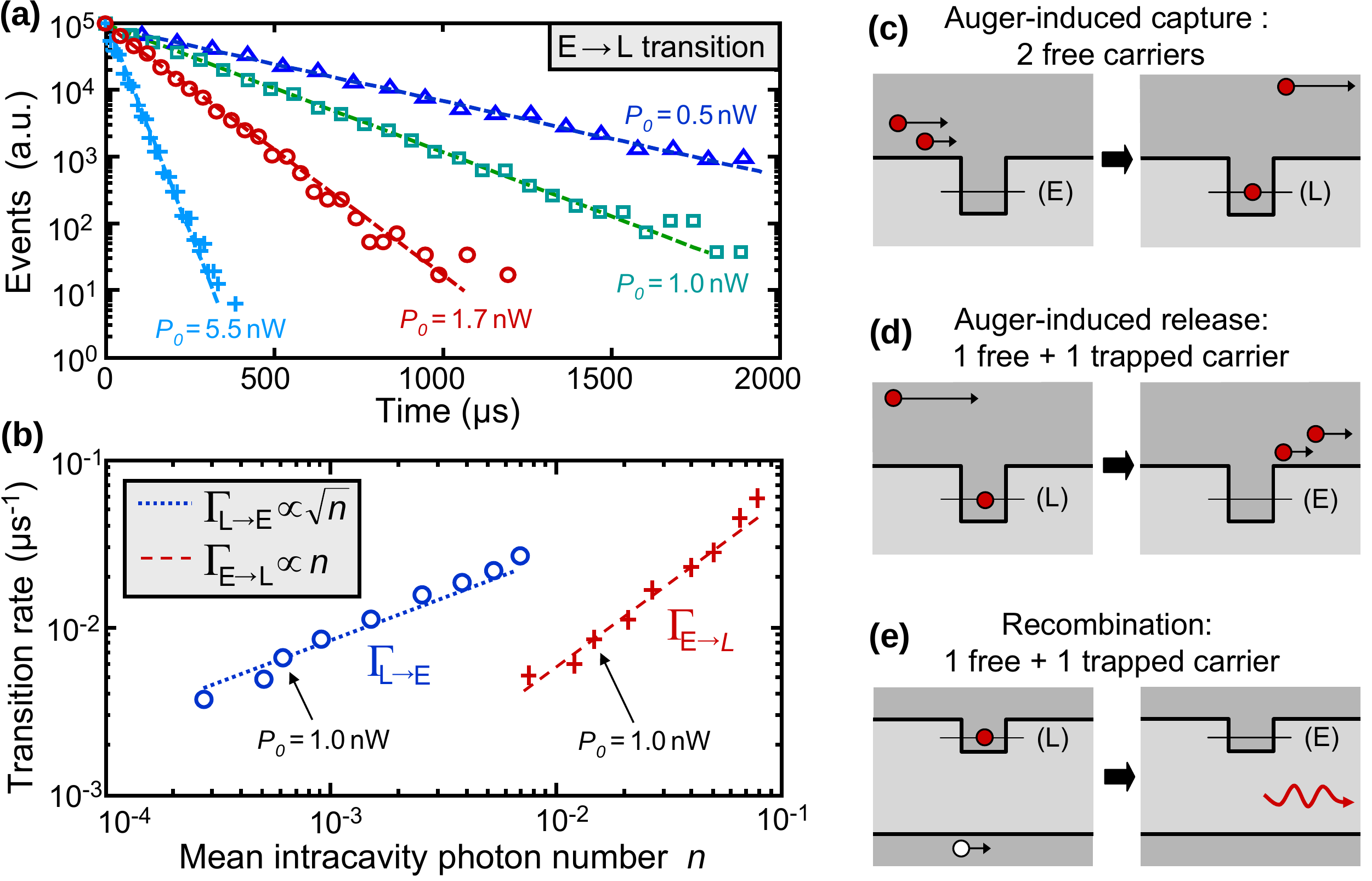}
\caption{\textbf{(a)} Histogram of the time spent in state $E$ before returning to state $L$, at $T=34.1~\mathrm{K}$ and $\omega=1.32411~\mathrm{eV}$, for various excitation powers. Dashed lines: numerical fits for monoexponential decays, from which the transition rates $\Gamma_{E \rightarrow L}$ are extracted. \textbf{(b)} Measured transition rates $\Gamma_{E \rightarrow L}$ and $\Gamma_{L \rightarrow E}$, extracted from the real-time measurements, as a function of the intracavity photon number $n$. Two different photon numbers are considered for each value of the incident power $P_0$: the photon number in the $E$ state governs the $\Gamma_{E \rightarrow L}$ transition rate, while the photon number in the $L$ state governs the $\Gamma_{L \rightarrow E}$ transition rate. Dashed and dotted lines: numerical fits (see legend). \textbf{(c)} to \textbf{(e)} Schematic view of various microscopic processes allowing for carrier capture and release events.  \label{Fig3}}
\end{figure}

The data in Fig. 3(b) allow one to deduce the empiric laws governing the increase of the transition rates with $n$. A satisfying fit of the experimental data is obtained with two different laws, namely:
\begin{equation} \label{eq_transition_rates}
\Gamma_{E \rightarrow L}  = \alpha \: n \ \ \ \ \textrm{and:} \ \ \ \ \  \Gamma_{L \rightarrow E} = \beta \:  \sqrt{n},
\end{equation}
where $\alpha=0.6~s^{-1}$ and $\beta=0.27 ~s^{-1}$ are empirical proportionnality coefficients. We propose a simple model to interpret these linear and sub-linear behaviors with the number of internal photons, considering that they participate to the generation of free carriers in the InAs wetting layer or in the GaAs barriers. Such a generation could be assisted by the interaction with phonons, with defects in the material, or with the other quantum dots in the microcavity. Following previous works \cite{Caroll1985,OHara1999,Berthelot2006,Sallen2011,Nguyen2013} we consider that the carrier generation rate is proportionnal to $n$, while the carrier recombination rate is proportionnal to $N_e N_h=N_c^2$, where $N_e=N_h$ are the electron and hole densities, considered both equal to the  carrier density denoted $N_c$. In this model where $\frac{\mathrm{d} N_c}{\mathrm{d} t}= a \: n - b \: N_c^2$, $a$ and $b$ being constant proportionnality coefficients, the stationnary regime $\frac{\mathrm{d} N_c}{\mathrm{d} t}=0$ corresponds to a carrier density $N_c$ proportionnal to $\sqrt{n}$. The empiric power laws deduced from our real-time measurements can thus be rewritten in the simple forms $\Gamma_{E \rightarrow L} \propto N_c^2$ and $\Gamma_{L \rightarrow E} \propto N_c$. 

As we now describe, such power laws find a direct interpretation within a simple model where capture and release events require the interaction between two carriers in the vicinity of the defect. For example, as illustrated in Fig. 3(c), an Auger capture process involves an inelastic collision between two free carriers nearby the material defect, one of which ends up captured in the trap: the expected interaction rate is then proportionnal to the square of the carrier density, in agreement with the power law $\Gamma_{E \rightarrow L} \propto N_c^2$. Conversely, as illustrated in Fig. 3(d), the release of the trapped carrier can also result from an Auger inelastic collision between the trapped carrier and an incoming free carrier; the expected interaction rate is proportionnal to the carrier density, in agreement with the linear law $\Gamma_{L \rightarrow E} \propto N_c$. Another physical process compatible with this law is when the interaction between the trapped carrier and an incoming free carrier leads to the recombination of both, as illustrated in Fig. 3(e); this again gives a transition rate $\Gamma_{L \rightarrow E} \propto N_c$. Similar models have been sucessfully invoked to interpret the experimental data in several different experiments \cite{Berthelot2006, Sallen2011, Nguyen2013}. However, none of these experiments were able to actually monitor in real-time the fluctuation events, and directly measure the capture and release transition rates, as is reported here.

We now show that the measured empiric laws of Eq. (1) are valid for a large range of experimental parameters, by analysing the real-time measurements performed as a function of the device temperature and laser wavelength. When tuning the temperature we observe a continuous change in the bare cavity mode energy $\omega_c$ at a rate of $17 \mathrm{µeV}/K$, and in the bare QD transition energies $\omega_{d}^{(L)}$ and $\omega_{d}^{(E)}$ at a rate of $81 \mathrm{µeV}/K$ \cite{Loo2010}, with a constant energy shift
$\omega_{d}^{(E)}-\omega_{d}^{(L)}=75~\mathrm{µeV}$. For each set of experimental conditions, 10 000 successive measurements are recorded, with a $2\: \mathrm{\mu s}$ integration time for each measurement and a total observation time of $20~\mathrm{ms}$; the experimental reflectivity average $R_{avg}$, and variance $\sigma_R^2$, are then extracted from these data.  Fig. 4(a) first displays in color scale the experimental average reflectivity, as a function of temperature and photon energy $\omega$, for an incident power $P_0=1.5~\mathrm{nW}$. Two anticrossings are observed with the cavity resonance, associated to the empty and loaded states, instead of only one anticrossing for a non-fluctuating device \cite{Loo2012}. 

\begin{figure}[ht!]
\centering \includegraphics[width=0.49\textwidth]{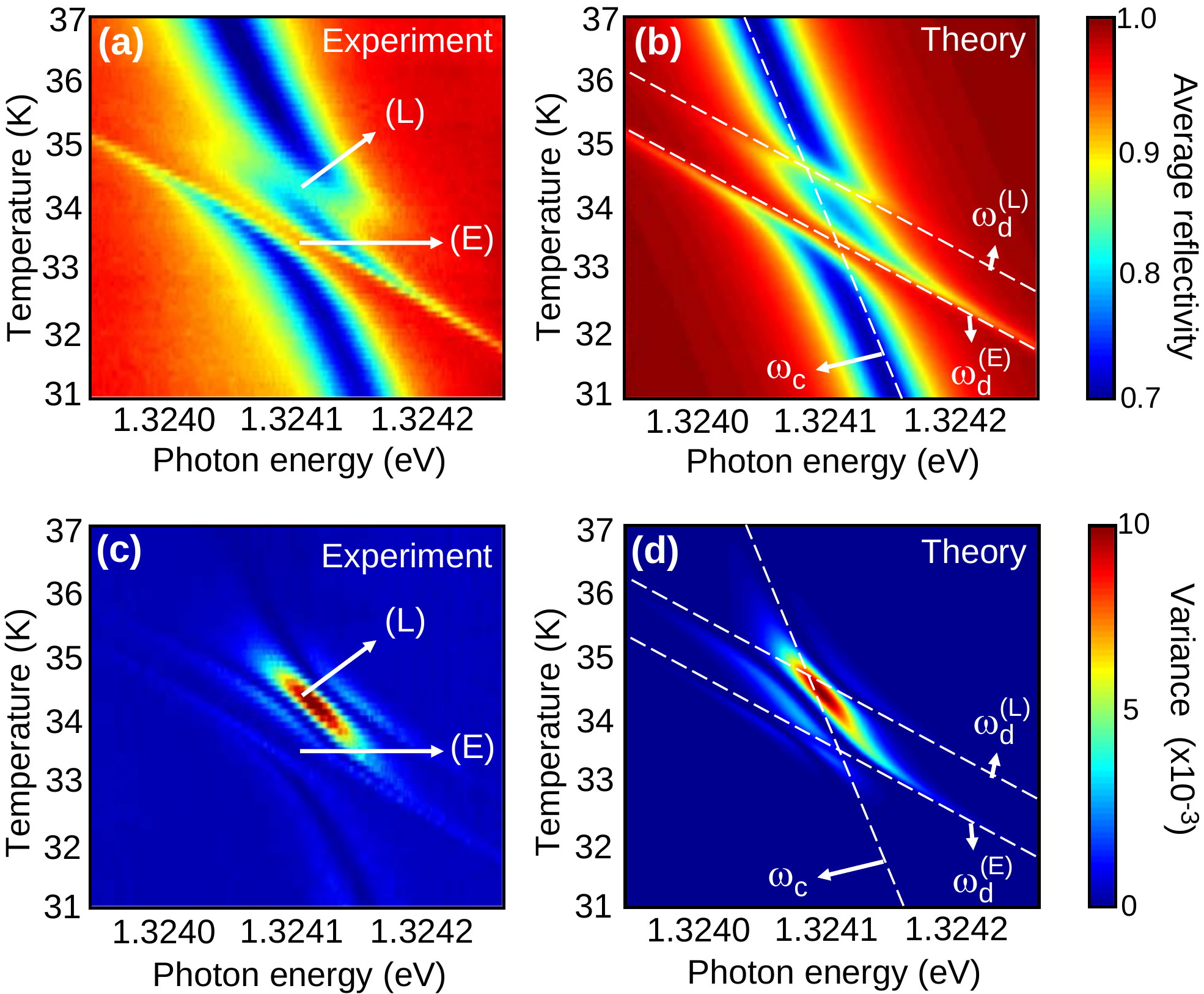}
\caption{\textbf{(a)} Experimental average reflectivity vs temperature and photon energy. \textbf{(b)} Simulated average reflectivity. Dashed lines: bare cavity mode and QD transition energies. \textbf{(c)} Experimental reflectivity variance vs temperature and photon energy. \textbf{(d)} Simulated reflectivity variance. 
 \label{Fig4}}
\end{figure}

To predict the theoretical average and variance of the reflectivity signal, we calculate the reflectivities for each state, $R_E$ and $R_L$, by applying the parameters already used in the fits of Fig. 2(a) without further adjustment. We also predict the intracavity photon numbers in the two states $E$ and $L$ \cite{SupportingInformation}; the theoretical transition rates $\Gamma_{E \rightarrow L}$ and $\Gamma_{L \rightarrow E}$ are then deduced using the power laws in Eq. (1). Then, the overall probabilities of the system being in the empty or loaded state, $P_E$ and $P_L$, are calculated using $P_E=\frac{\Gamma_{L \rightarrow E}}{\Gamma_{L \rightarrow E}+\Gamma_{E \rightarrow L}}$ and $P_L=1-P_E$. The theoretical average reflectivity  $R_{avg}=P_E R_E + P_L R_L$ is plotted in Fig. 4(b), showing a good agreement with the experimental data. As a guide to the eyes the values of the bare cavity mode frequency $\omega_c$, and of the bare QD transition frequencies $\omega_{d}^{(E)}$ and $\omega_{d}^{(L)}$, have been indicated: the anticrossings occur for $\omega_c \approx \omega_{d}^{(E)}$ and  $\omega_c \approx \omega_{d}^{(L)}$. 


A fair agreement is also obtained between the experimental and calculated variance colormaps, respectively displayed in Fig. 4(c) and 4(d). In the latter case the theoretical variance is given by $\sigma_R^2=P_E P_L(R_E-R_L)^2$: a high variance $\sigma_R^2$ thus comes from different reflectivities $R_E$ and $R_L$ but also comparable probabilities $P_E$ and $P_L$. Only one region of high variance is obtained, around the loaded-state anticrossing where there is a maximal reflectivity difference $R_L-R_E$. In this specific region where $R_L$ is close to unity, the intracavity photon number is much lower when the system is in the loaded state, most of the incident photons being reflected. The power laws of Eq. (1) then lead to approximately equal transition rates $\Gamma_{L \rightarrow E}$ and $\Gamma_{E \rightarrow L}$, thus to probabilities $P_E$ and $P_L$ close to $\frac{1}{2}$. This region of high variance corresponds to the one highlighted in Fig. 2(a), which is where the real-time monitoring of Fig. 2(b) has been performed; the approximate equality $P_E\approx P_L\approx \frac{1}{2}$ is indeed verified in the histogram of Fig. 2(c). Almost no variance is observed, on the contrary, near the empty-state anticrossing where the difference $R_E-R_L$ is also high. In this region, $R_E$ is close to unity and the photon number is much lower when the system is in the empty state. The power laws of Eq. (1) then imply a release rate $\Gamma_{L \rightarrow E}$ much lower than the capture rate $\Gamma_{E \rightarrow L}$, leading to $P_L << P_E$ and a negligible variance; this is the configuration also occuring in the low-variance region highlighted in Fig. 2(a). Finally, the same arguments allow us to understand why, in figures 4(a) and 4(b), the loaded-state anticrossing is less contrasted than the empty-state anticrossing. The first case indeed corresponds to $P_E \approx P_L$ and thus $R_{avg}\approx \frac{R_E+R_L}{2}$, while the second one corresponds to $P_L << P_E$ and thus $R_{avg}\approx R_E$.

To conclude, we have shown that coherent reflectivity measurements allow the real-time monitoring of single-charge jumps at the microsecond timescale. The capture and release of a charge by a material defect has been directly observed in a strongly-coupled QD-cavity system, highly sensitive to single-carrier fluctuations in the QD environment. All the experimental observations allow to propose an empiric model, where the fluctuation processes are governed by the interaction between two carriers in the vicinity of the defect. Our results also evidence the back-action of the measurement on the system dynamics, as free carriers are generated in the wetting layer or in the GaAs barriers in spite of the purely resonant excitation scheme. Fast coherent measurements, beyond providing a novel tool in the harnessing of solid-state fluctuations, can also be extended to the real-time monitoring of a single electron or hole spin in a charged quantum dot. This would constitute a quantum non-demolition experiment, one of the fundamental building blocks of a spin-photon interface, where the spin state is projected by the measurement onto one of its two possible eigenstates.

We would like to thank H.S. Nguyen and I. Favero for fruitful discussions. This work was partially supported by the French ANR MIND, ANR CAFE, ANR QDOM, the ERC starting grant 277885 QD-CQED, the CHISTERA project SSQN, and by the French RENATECH network.


\providecommand*\mcitethebibliography{\thebibliography}
\csname @ifundefined\endcsname{endmcitethebibliography}
  {\let\endmcitethebibliography\endthebibliography}{}

\end{document}